\documentclass[dvips,12pt]{article}
\usepackage{amsmath,amssymb,exscale}   
\usepackage{array,multicol}
\usepackage{afterpage,float,flafter}   
\usepackage{epsfig,rotating,pifont,fancybox}   
\makeatletter   
\@addtoreset{equation}{section}   
\makeatother   
   
\def\simgt{\stackrel{>}{{}_\sim}}
\newcommand{\bea}{\begin{eqnarray}}   
\newcommand{\eea}{\end{eqnarray}}   
   
\newcommand{\NPB}[3]{\emph{ Nucl.~Phys.} \textbf{B#1} (#2) #3}   
\newcommand{\PLB}[3]{\emph{ Phys.~Lett.} \textbf{B#1} (#2) #3}   
\newcommand{\PRD}[3]{\emph{ Phys.~Rev.} \textbf{D#1} (#2) #3}   
\newcommand{\PRL}[3]{\emph{ Phys.~Rev.~Lett.} \textbf{#1} (#2) #3}

\newcommand{\RMP}[3]{\emph{ Rev.~Mod.~Phys.} \textbf{#1} (#2) #3}   
\newcommand{\HPA}[3]{\emph{ Helv.~Phys.~Acta} \textbf{#1} (#2) #3}

\newcommand{\JHEP}[3]{\emph{ JHEP} \textbf{#1} (#2) #3}  
\newcommand{\Tr}{\mathop{\rm Tr}}

\title{   
\vspace*{-1.3cm}   
\begin{flushright}   
\normalsize{      
IEM-FT-225/02\\
FERMILAB-PUB-02/159-T \\
ANL-HEP-PR-02-041 \\
EFI-2002-87  \\
\textsf{hep-ph/0208043}
}
\end{flushright}    
\vspace{.5cm}
\Large {\sc Improved Results in Supersymmetric Electroweak
Baryogenesis~\footnote{Work 
supported in part by CICYT, Spain, under
contract FPA2001-1806, and by EU under contracts HPRN-CT-2000-00152
and HPRN-CT-2000-00148.}}
\vspace*{.2cm}
\author{\large\sc
{M.~Carena~$^a$, M.~Quir\'os~$^{b}$,
M.~Seco~$^{c}$ and C.E.M.~Wagner~$^{d,e}$}\\ \\
$^a$\normalsize\emph{Fermi National Accelerator Laboratory,
P.O. Box 500, Batavia, IL 60510, USA }\\
$^b$\normalsize\emph{Instituto de Estructura de la Materia (CSIC),
Serrano 123, 
E-28006 Madrid, Spain}\\
$^c$\normalsize\emph{Dep. of Phys., 
U. of Virginia, 382 McCormick Road, 
Charlottesville, VA 22904-4714, USA
}\\
$^d$\normalsize\emph{HEP Division, Argonne National Laboratory,
9700 Cass Ave,
Argonne, IL 60439, USA} \\
$^e$\normalsize\emph{Enrico Fermi Institute, Univ. of Chicago, 5640
Ellis Ave., Chicago, IL 60637, USA}}}
\date{}   
\begin{document}
\maketitle


\begin{abstract}
Electroweak baryogenesis provides a very attractive scenario to
explain the origin of the baryon asymmetry. The mechanism of
electroweak baryogenesis makes use of the baryon number anomaly and
relies on physics that can be tested experimentally. It is today
understood that, if the Higgs mass is not larger than 120 GeV, this
mechanism may be effective within supersymmetric extensions of the
Standard Model. In this work, we reconsider the question of baryon
number generation at the electroweak phase transition within the
context of the minimal supersymmetric extension of the Standard
Model. We derive the relevant diffusion equations, give a consistent
definition of the sources, and compare our results with those
appearing in the recent literature on this subject.
\end{abstract}

\thispagestyle{empty}
\newpage

\section{\sc Introduction} 
\label{introduction}

Electroweak Baryogenesis~\cite{reviews} provides a predictive
framework for the computation of the baryon asymmetry of the
Universe~\cite{baryogenesis}. Perhaps the most attractive feature of
this mechanism is that it relies on anomalous baryon number violation
processes which are present in the Standard Model~\cite{anomaly}. At
temperatures far above the electroweak phase transition scale, these
anomalous processes are unsuppressed and, in the absence of any $B-L$
asymmetry, they lead to the erasure of any baryon or lepton number
generated at high energy scales~\cite{sphalT}. These baryon number
violation processes are, instead, exponentially suppressed in the
electroweak symmetry broken phase, at temperatures far below the
electroweak phase transition~\cite{sphalerons}. At the electroweak
phase transition, non-equilibrium processes may generate a
non-vanishing baryon number which may efficiently diffuse into the
broken phase~\cite{ckn}. The generated baryon number depends on the
CP-violating currents present in the model.  The mechanism of
electroweak baryogenesis may become effective if the CP-violating
sources are strong and, at the electroweak phase transition
temperature, the baryon number violation processes in the broken phase
are sufficiently suppressed, leading to a baryon number density in the
broken phase consistent with observations. This, in turn, demands a
strongly first order electroweak phase transition.

It has been long ago realized that in the Standard Model the CP
violating sources are too weak to lead to an acceptable baryon number
density~\cite{fs,huet}. Moreover, even if the sources were strong
enough to lead to a reasonable baryon number generation, the phase
transition is weakly first order, leading to a strong exponential
suppression of the baryon number generated in the broken
phase~\cite{SMpt}. A minimal supersymmetric extension of the Standard
Model (MSSM), instead, has all the necessary ingredients to improve on
both problems~\cite{early}. First, there are additional sources of
CP-violation, associated with the CP-violating phases of the
supersymmetry breaking parameters. Second, in the presence of a light
stop, the phase transition may become strong enough to allow the
preservation of the baryon number generated at the electroweak
symmetry breaking scale~\cite{mariano1}-\cite{LR}.

Although there is general agreement on both the existence and nature
of the new sources and also on the parameter space leading to a
strongly first order phase transition, the results regarding the
strength and specific form of the CP-violating sources are still
controversial. Most groups working on the subject have found that, for
values of the CP-odd Higgs mass of the order of the weak scale, the
sources are dominated by a term proportional to $\epsilon_{ij} H_i
\partial^{\mu} H_j$, where $H_i$ denote the expectation value of the
neutral components of the two Higgs doublets present at low
energies~\cite{hn}-\cite{wefive}. Recently, however, it has been
claimed that this contribution is absent, leading to a suppressed
result for the baryon asymmetry contribution within the MSSM
proportional to $H_1 \partial^{\mu} H_2 + (1 \leftrightarrow
2)$~\cite{plus,Kainulainen:2001cn}.

In this article, we proceed to perform a critical re-analysis of the
diffusion equations leading to the baryon asymmetry and of the sources
for those diffusion equations. In particular, we demonstrate that the
sources can be expressed in terms of appropriate differential
operators times the CP-violating currents we computed in a very recent
work. We find no suppression of the dominant sources proportional to
$\epsilon_{ij} H_i \partial^{\mu} H_j$. Indeed, we show that the
suppression claimed the authors of
Refs.~\cite{plus,Kainulainen:2001cn} is an artifact of the
approximation they used in order to compute the currents.

The organization of the article is as follows. In section 2 we provide
the derivation of the diffusion equations. In section 3 we review the
derivation of the CP-violating currents and we show explicitly where
our results differ from the ones of Ref.~\cite{plus}. In section 4 we
discuss the results for the baryon asymmetry within the MSSM and
section 5 contains our conclusions. Finally some useful formulae
concerning the chargino sector of the MSSM are summarized in
appendix~A.

\section{\sc Diffusion equations}
\label{diffusioneq}

We will start considering a system of particles propagating in a
non-trivial CP-violating background localized in the bubble wall,
where the bubble is expanding with a speed $v_\omega$ in the plasma
frame.  The presence of the CP-violating background and the particle
number changing reactions perturb the distribution functions for
particles and antiparticles around the equilibrium.  The corresponding
distribution functions $f_i$ satisfy the Boltzmann equations,
\begin{equation}
\label{boltzmann0}
v^\mu\partial_\mu\, f_i+
\bar{F}^\mu_i\,\nabla_\mu\, f_i=C_i[f]
\end{equation} 
where we are using the notation $\partial_\mu\equiv\partial/\partial
x^\mu$, $\nabla_\mu\equiv\partial/\partial p^\mu$, $v^\mu=d
x^\mu/dt\equiv p^\mu/E$ is the four-velocity and
$\bar{F}^\mu=dp^\mu/dt$ is the force generated by the non-trivial
background. The term on the right-hand side $C_i=\left(\partial
f_i/\partial t \right)_{\rm coll}$ encodes the effect on the
distribution of the particle number changing reactions and elastic
collisions.

In the bubble-wall frame and neglecting the curvature of the bubble wall the
distribution functions can be written as,
\begin{equation}
\label{fwall}
f^P_i=\frac{\bar{g}(p_z,z)}{e^{\beta[E+v_\omega p_z-\mu_i^P]}\pm
1}+\delta f_i^P
\end{equation}
where $\mu_i^P=\mu_i^P(z)$ is the chemical potential, $E$ and $p_z$
are the energy and momentum of the particle, the momentum-dependent
part $\delta f_i^P =\delta f_i^P (E,p_z;z)$ describes departure of the
system from kinetic equilibrium and we have introduced a slowly
varying CP-even function of $p_z$ and $z$, $\bar{g}(p_z,z)$, which
tends to 1 in the absence of forces. Considering the effective
distribution functions $f_i = f^P_i - f^{\bar{P}}_i$, describing the
difference between particles and anti-particles of a given species,
and applying (\ref{fwall}) to Eq.~(\ref{boltzmann0}) gives
 \begin{equation}
\label{boltzwall}
\frac{p_z}{E} \left[-\mu'_i\, \bar{g}(p_z,z)
\frac{\partial f_0}{\partial E}+\delta f'_i \right]+
\left(F_i^0+v_\omega F_i^z \right) \bar{g}(p_z,z)
\frac{\partial f_0}{\partial E}=C_i
\end{equation}
where $F_i^{\mu}$ is the CP-violating component of the force, we have
linearized with respect to the small perturbations, $\mu_i$, $\delta
f_i$, in the spirit of the Chapman-Enskog 
first order approximation~\cite{Chapman},
$f_0$ is the equilibrium distribution function in the plasma frame,
and the $(')$ denotes derivative with respect to $z$. In this process,
due to the assumed smoothness of the CP-even function
$\bar{g}(p_z,z)$, we have ignored terms proportional to derivatives of
this function.

Following Ref.~\cite{plus}, we rewrite the above equation in the
plasma frame by performing a Galilean transformation on the velocity
$v_z$ to $v_z - v_\omega$,
\begin{equation}
\label{boltzplasma}
\left(\frac{p_z}{E} - v_\omega \right) \left[-\mu'_i\, g(p_z,z) \;  
\frac{\partial f_0}{\partial E}+\delta f'_i \right]+
\left(F_i^0+v_\omega F_i^z \right) g(p_z,z)
\frac{\partial f_0}{\partial E}=C_i
\end{equation}
where $g(p_z,z)$ is assumed to be an even function of momenta in this
frame.

We now define
\begin{equation}
\label{definicion}
\langle X \rangle\equiv \int \frac{d^3 p}{(2\pi)^3}\,X,
\end{equation}
and using the decomposition 
\begin{equation}
\label{decomp}
f_i= -\mu_i \frac{\partial f_0}{\partial E} g(p_z,z)
+\delta f_i
\end{equation}
and the definition of the current
\begin{equation}
\label{corriente}
j_i^\mu\equiv\langle v^\mu f_i \rangle=\left( n_i,j_i^z \right)
\end{equation}
we can relate the CP-violating current with the chemical potential and
the function $\delta f_i$ as,
\begin{align}
\label{calculos}
n'_i=& -\left< \frac{\partial f_0}{\partial E} g(p_z,z) \right>
\mu'_i\nonumber\\ j_i^z =& \left< \frac{p_z}{E}\, \delta f_i \right>,
\end{align}
where, again, we have neglected terms proportional to derivatives of
$g(p_z,z)$. The function $\delta f_i$ is such that $\left<\delta f_i
\right>=0$ and the number density is entirely provided by the chemical
potential $\mu_i$. In this way we can choose, with all generality, the
function $\delta f_i$ as odd with respect to $v_z$ and therefore
satisfying the property $\left<v_z^{2n}\delta f_i \right>=0$.

We now multiply Eq.~(\ref{boltzplasma}) by $v^\mu=(1,p_z/E)$ and 
integrating over momenta we get the two equations,
\begin{align}
\label{ecua1}
-v_\omega n'_i+\left(j_i^z \right)'+\left< \left(F_i^0+v_\omega F_i^z\right)
g(p_z,z)
\frac{\partial f_0}{\partial E}\right>=&\left<C_i \right>\\
D_i \Gamma_i^T n'_i-v_\omega\left(j_i^z \right)'+
\left< \frac{p_z}{E}\,\left(F_i^0+v_\omega F_i^z\right)
g(p_z,z)
\frac{\partial f_0}{\partial E}\right>=&\left<\frac{p_z}{E}\,C_i \right>
\label{ecua2}
\end{align}
where $\Gamma_i^T$ is the total interaction rate and
the diffusion coefficient $D_i$ is defined as
\begin{equation}
\label{diffusion}
D_i=\frac{1}{\Gamma_i^T}\,\frac{\left<{\displaystyle \frac{p_z^2}{E^2} 
\frac{\partial f_0}{\partial E}}\right>}{
\left<{\displaystyle \frac{\partial f_0}{\partial E}} \right>} 
\end{equation}
and we have again ignored the smooth momentum dependence of $g(p_z,z)$
in the numerator and denominator integral functions. Observe that the
ratio (\ref{diffusion}) remains constant up to derivatives of
$g'(p_z,z)$ which we have consistently ignored in our treatment.

The terms on the right-hand sides of Eqs.~(\ref{ecua1}) and (\ref{ecua2}) 
can be decomposed, using the decomposition (\ref{decomp}) as
\begin{equation}
\label{c}
C_i[f]=C_i^1[\mu]+C_i^2[\delta f]
\end{equation}
where $C_i^1$ and $C_i^2$ are linear in $\mu_i$ and $\delta f_i$,
respectively.  In fact $\left<C_i^1[\mu]\right>$ is entirely provided
by inelastic collisions
\begin{equation}
\label{c1}
\left<C_i^1[\mu]\right>\simeq -\Gamma_{ij} \frac{n_j}{k_j}
\end{equation}
where $k_j=2$ ($k_j=1$) for bosons (fermions) are statistical factors,
$\Gamma_{ij}$ ($j=j_1,j_2,\dots$) is the averaged interaction rate for
the inelastic reaction channel ($i,j$) and $n_j/k_j$ implies a sign
over all particles participating in the reaction with appropriate sign
assignments: For an inelastic reaction corresponding to the decay
$i\to j_1\, j_2$ with decay width $\Gamma_{ij_1j_2}$, its contribution
to the right-hand side of (\ref{c1}) would be
$-\Gamma_{ij_1j_2}(n_i/k_i-n_{j_1}/k_{j_1}-n_{j_2}/k_{j_2})$.
Furthermore $\left<C_i^2[\delta f]\right>=0$ from the oddness of
$\delta f_i$.  On the other hand, $\left<v_z\,C_i^1[\mu]\right>=0$.

Within the present framework, the CP-violating forces are provided by
the interaction of the different fields with the Higgs profiles, which
vary along the bubble wall. These CP-violating forces induce
CP-violating currents for the different fields, which lead to
$(n_i,j_i^z)$ after elastic and inelastic interactions with the other
fields present in the plasma. The resulting currents have then two
components. The first one, $(n_i^{(B)},j_i^{(B)})$, the background
component, is the one that would be obtained in the presence of only
the interaction with the background Higgs field, and the second one,
the ``collision'' component comes from the interaction with the plasma
fields and lead to the diffusion process.  The integral
$\left<v_z\,C_i^2[\delta f]\right>$ is contributed by all collisions
and dominated by the elastic ones. Correspondingly with the two
components, background and collision, of the currents there can be
defined two $\delta f$ components.

 The part of $\delta f$ that is induced by the interactions with the
plasma fields, the collision component, is governed by the same
interactions as the ones governing $C_i$ and is therefore relevant in
the same regime of momenta in which the collision term becomes
important. On the other hand the $\delta f$ component coming from the
interaction with the background fields, $\delta f^{(B)}$, is not
correlated with the collision terms and leads naturally to a
negligible contribution to $\left<\frac{p_z}{E}\,C_i^2[\delta
f]\right>$.  Therefore, considering that the rate of interactions is
approximately constant in the momentum regime in which the collision
contribution to the currents become relevant, and negligible anywhere
else, we get
\begin{equation}
\label{c2}
\left<\frac{p_z}{E}\,C_i^2[\delta f]\right>
\simeq \Gamma_i^T\, \left(j_i^z - j_i^{(B)z} \right) .
\end{equation}

Using now (\ref{c1}) and (\ref{c2}) in (\ref{ecua1}) and (\ref{ecua2}) we 
obtain the diffusion equations,
\begin{align}
\label{ecua3}
-v_\omega n'_i+\left(j_i^z \right)'+\left< \left(F_i^0+v_\omega
F_i^z\right) g(p_z,z) \frac{\partial f_0}{\partial
E}\right>=&-\Gamma_{i,j} \frac{n_j}{k_j}\\ D_i \Gamma_i^T
n'_i-v_\omega\left(j_i^z \right)'+ \left<
\frac{p_z}{E}\,\left(F_i^0+v_\omega F_i^z\right) g(p_z,z)
\frac{\partial f_0}{\partial E}\right>=&\Gamma_i^T\, \left(j_i^z -
j_i^{(B)z}\right).
\label{ecua4}
\end{align}

A current $j_i^{(B)\,\mu}=\left(n_i^{(B)},\, j_i^{(B)\, z}\right)$ in
the presence of a CP-violating Higgs background (or CP-violating force
$F^\mu$) but where no interactions with the plasma fields were
considered was computed in Ref.~\cite{wefive}. It should satisfy the
set of Boltzmann equations
\begin{align}
\label{ecua3B}
-v_\omega \left(n_i^{(B)}\right)'+\left(j_i^{(B)\,z} \right)'
+\left< \left(F_i^0+v_\omega F_i^z\right)
g(p_z,z)
\frac{\partial f_0}{\partial E}\right>=&0\\
D_i \Gamma_i^T \left(n_i^{(B)}\right)'-v_\omega\left(j_i^{(B)\, z} \right)'+
\left< \frac{p_z}{E}\,\left(F_i^0+v_\omega F_i^z\right)
g(p_z,z)
\frac{\partial f_0}{\partial E}\right>=& 0
\label{ecua4B}
\end{align}
We can see from these equations, and in particular from (\ref{ecua3B})
that the roles played by the CP-violating forces and currents
$(n_i^{(B)},\, j_i^{(B)})$ are equivalent. In fact by subtracting
(\ref{ecua3B}) from (\ref{ecua3}) one obtains an equation
\begin{equation}
\label{ecua3f}
-v_\omega \left(n_i-n_i^{(B)}\right)'+\left(j_i^z-j_i^{(B)\,z} \right)'
+\Gamma_{ij} \frac{n_j}{k_j}=0
\end{equation}
where the force terms have been replaced by the divergence of the CP-violating
current.

By subtracting now (\ref{ecua4B}) from (\ref{ecua4}) we obtain,
\begin{equation}
\label{fick}
\left(j_i^z-j_i^{(B)\,z} \right)=D_i\left[ \left(n_i-n_i^{(B)}\right)'
-\frac{v_\omega}{\Gamma_i^T}\left(n_i-n_i^{(B)}\right)''+\mathcal{O}(v_w^2)
\right]\simeq D_i \left(n_i-n_i^{(B)}\right)'
\end{equation}
where the last approximation comes from the requirement
$v_\omega/\Gamma_i^T L_w\ll 1$ (where $L_w$ is the wall thickness), 
which is
necessary for the validity of the derivative expansion we will use in the
calculation of the CP-violating currents in section~\ref{sources}. 
Eq.~(\ref{fick}) represents the well-known empirical law of diffusion or
Fick's law.

Replacing now Eq.~(\ref{fick}) into (\ref{ecua3f}) we obtain the final 
expression for the diffusion equation as,
\begin{equation}
\label{ecfinal}
-v_\omega n'_i+D_i n''_i+\Gamma_{ij}\frac{n_j}{k_j}=S_i[n^{(B)}]
\end{equation}
where the source is given by
\begin{equation}
\label{source}
S_i=D_i\left(n_i^{(B)} \right)''-v_\omega \left(n_i^{(B)} \right)'
\simeq D_i\, \left(n_i^{(B)} \right)''
\end{equation}
and the last approximation comes from the fact that $D_i\gg L_\omega\,
v_\omega$. Notice that Eq.~(\ref{ecfinal}) has the correct boundary
conditions since in the absence of inelastic reactions (for
$\Gamma_{ij}=0$) it provides the trivial solution $n_i=n_i^{(B)}$ as
required.

We can compare this result with the source obtained in
Ref.~\cite{plus}. These authors work in a WKB approximation with
quasi-particles. Within their semiclassical approximation the energy
is a (constant) label of WKB-states and therefore the time component
of the force $F_i^0=dE/dt$ vanishes, while they have assumed that
$g(p_z,z) = 1$.  Moreover their source, $S_i^{CJK}$, is given
by~\cite{plus},
\begin{equation}
\label{sourceCJK}
S_i^{CJK}=-\frac{1}{\Gamma_i^T}\,
\left< \frac{p_z}{E}\,v_\omega F_i^z\,
\frac{\partial f_0}{\partial E}\right>'
\end{equation}
which appears in Eq.~(\ref{ecua4B}) above and can therefore be related
to $n_i^{(B)}$ and $j_i^{(B)\, z}$ as they also appear in
(\ref{ecua4B}).  Although we have obtained the sources $S_i$ in a way
independent of the form of the CP-violating forces $F_i^z$ and
$F_i^0$, for vanishing $F_i^0$ we obtain a relation
between the sources $S_i$ and $S_i^{CJK}$. Indeed, using
Eq. (\ref{ecua4B}), we obtain
\begin{equation}
\label{comp}
S_i^{CJK}\simeq D_i\left(n_i^{(B)} \right)''-\frac{v_\omega}{\Gamma_i^T}
\left(j_i^{(B)\, z}\right)'
\simeq S_i .
\end{equation}
where the last approximation relies on the relation $D_i \Gamma_i^T\gg 1$. 
The latter relation holds for the case of charginos in the MSSM, analyzed in 
section~\ref{sources}.

\section{\sc The sources}
\label{sources}

In this section we will make contact between the source in
(\ref{source}), and in particular the current $j^{(B)\,\mu}$, and the
Green function and will apply the formalism to the sector of charginos
in the MSSM.

First consider a chiral fermion (say a right-handed one) in the
presence of the non-trivial background.  Its Green function
$S^{(B)}(x,y)$ can be considered as function of the center-of-mass
coordinate $z=(x+y)/2$ and the relative coordinate $r=x-y$,
$$S^{(B)}(z+r/2,z-r/2).$$ Since we will make a gradient expansion with
respect to the coordinate $z$ we perform a Fourier transformation with
respect to the relative coordinate $r$ as
\begin{equation}
\label{Wigner}
S^{(B)}(z;p)=\int d^4r\ 
e^{ip\cdot r} \
S^{(B)}(z+r/2,z-r/2)
\end{equation}
which is called the Wigner representation of the Green function $S^{(B)}$.

By making the general decomposition
\begin{equation}
\label{decompS}
S^{(B)}=\sigma^\mu S^{(B)}_\mu(z;p)
\end{equation}
one can define the corresponding background current $j^{(B)\,\mu}$ as
\begin{equation}
\label{cor}
j^{(B)\,\mu}=\int\frac{d^4p}{(2\pi)^4}\ S^{(B)\,\mu}
\end{equation}
and making the inverse Wigner transformation (\ref{Wigner}) it can be cast as
\begin{equation}
\label{cor2}
j^{(B)\,\mu}=\frac{1}{2}\lim_{r\to 0}\Tr \sigma^\mu S^{(B)}(z+r/2,z-r/2)
\end{equation}

The current can be written in an alternative equivalent form, that
makes connection with the discussion in the previous section,
\begin{equation}
j^{(B) \; \mu} = - 
\int \frac{d^4p}{(2 \pi)^4} \; p^{\mu} \, \nabla^\nu S^{(B)}_\nu(z;p)
= \int \frac{d^4p}{(2 \pi)^4} \; S^{(B)}_\mu
\end{equation}
Under small perturbations the relevant component of the integrand is
approximately dominated by $\delta$ functions imposing the dispersion
relation between $p_0$ and the energy of the particle. Indeed, for a
free particle in the presence of a non-trivial chemical potential,
the Green functions (\ref{relations}) and (\ref{prop}) lead to
\begin{equation}
S_{\mu} =  i  \frac{p_{\mu}}{p^2 - m^2 + i \epsilon} + 
          2  \pi \; p_{\mu} \; f(p_0,\mu) \;   
\delta(p^2 - m^2) ,
\end{equation}
where $p^2 = p^{\mu} p_{\mu}$ and, since the momentum integral of the
first term vanishes, the charge density is obtained from the
integration of the second term. Under these conditions,
Eqs.~(\ref{cor}) and (\ref{cor2}) are consistent with the definitions
of the particle density $n^{(B)}$ and current $\vec{j}^{(B)}$ in
(\ref{corriente}).
%
%

For fermions of opposite chirality (say left-handed ones) the previous
expressions hold just changing $\sigma^\mu \to \bar\sigma^\mu$. On the
other hand for several flavors, the mass eigenstates are an admixture
of the weak eigenstates, with a mixing that depends on the value of
the varying Higgs background.  The expression of the diffusion
equations and the sources, Eqs.~(\ref{ecfinal}) and (\ref{source}),
are related in a very simple way; namely the sources are trivially
obtained by demanding a self-consistency relation in the absence of
particle changing interactions.  This property should hold, in first
approximation, for any flavor structure of the theory. The relevant
sources for the diffusion equations should be obtained by generalizing
the trace in (\ref{cor2}) to flavor space and eventually including a
projection operator inside the trace, $\mathcal{P}$, when a given
contribution is to be picked up.  As stressed above, this formalism
makes contact with the one proposed earlier by the authors of
Ref.~\cite{plus} and should lead to a more realistic estimate of the
baryon asymmetry of the universe than the one presented in previous
analyses of this subject.

Although, as we have already outlined, we agree with the results of
Ref.~\cite{plus} in the formal definition of the sources, we shall now
show that the structure of these sources differs from the one
presented by the authors of Ref.~\cite{plus}.  We shall try to clarify
the origin of the discrepancy with those authors' results.

For the chargino sector of the MSSM we define the right-handed and 
left-handed fermions as
\begin{align}
\psi_R(x)=
\begin{pmatrix}
\widetilde{W}^+\\\widetilde{h}_2^+
\end{pmatrix},
&\qquad
\psi_L(x)=
\begin{pmatrix}
\widetilde{W}^-\\\widetilde{h}_1^-
\end{pmatrix}\notag \ .
\end{align}
and expand the mass matrix to first order in derivatives around the
point $z$ as
\begin{equation}
\label{Mexp}
M(x)=M(z)+(x-z)^\mu M_\mu(z) \ ,
\end{equation}
where we use the notation $M_\mu(z)\equiv\partial M(z)/\partial z^\mu$, and 
the mass matrix $M(z)$ is defined in (\ref{masach}).

We now consider the first term in (\ref{Mexp}) as part of the unperturbed 
Lagrangian, and the second term as the perturbation
\begin{equation}
\label{splitlag}
\mathcal{L}_{int}(x)=(x-z)^\mu\left\{
\psi_R^{\dagger}(x)M_\mu(z)\psi_L(x)+
\psi_L^{\dagger}(x)M_\mu^{\dagger}(z)\psi_R(x)\right\} \ .
\end{equation}
or, in the basis of mass eigenstates,
\begin{align}
\chi_R(x)=\mathcal{U}(z)\psi_R(x),&\qquad
\chi_L(x)=\mathcal{V}(z)\psi_L(x)\notag
\end{align}
\begin{equation}
\label{philag}
\mathcal{L}_{int}(x)=(x-z)^\mu \left\{
\chi_R^{\dagger}(x)\,\mathcal{U}(z) M_\mu(z)
\mathcal{V}^{\dagger}(z)
\chi_L(x)+\chi_L^{\dagger}(x)\mathcal{V}(z) 
M^{\dagger}_\mu(z)\,\mathcal{U}^{\dagger}(z)
\chi_R(x) \right\}
\end{equation}

We denote by $S^{LL}$, $S^{RR}$, $S^{LR}$ and $S^{RL}$ the left-left,
right-right, left-right and right-left Green functions of free
fermions in the mass eigenstate basis with mass $m_i(z)$ as given in
(\ref{relations}). In this basis the right- and left-handed Green
functions get modified to $S_\chi^{RR}$ and $S_\chi^{LL}$ by the
presence of the interaction term (\ref{philag}) and can be defined as
solutions of Schwinger-Dyson equations. To first order in the
insertion given by the interaction term (\ref{philag}) the solution to
the Schwinger-Dyson equations is
\begin{align}
\label{cgreen}
S_\chi^{RR}(x,y;z)=&S^{RR}(x,y;z)+\Delta S_\chi^{RR}(x,y;z)\notag\\
\Delta S_\chi^{RR}(x,y;z)=&\int{d^4w(w-z)^\mu 
S^{RR}(x,w;z)\,\mathcal{U}(z)\, M_{\mu}(z)
\mathcal{V}^{\dagger}(z)\,S^{LR}(w,y;z)}\notag\\
+&\int{d^4w(w-z)^\mu S^{RL}(x,w;z)\mathcal{V}(z)\,
M^{\dagger}_{\mu}(z)\mathcal{U}^{\dagger}(z)S^{RR}(w,y;z)}
\end{align}
and a similar solution for $S_\chi^{LL}$.

The CP-violating Green function $\Delta S_\chi^{RR}(x,y;z)$ defined in
(\ref{cgreen}) can be written as,
\begin{equation}
\label{inicio}
\Delta S_\chi^{RR}=\frac{1}{4}\int \frac{d^4 p}{(2\pi)^4}\,
e^{i\,p\cdot r}
\left\{\sigma^\mu G\, \mathcal{U}\, M_\mu\, M^\dagger\, 
\mathcal{U}^\dagger\,
G+\sigma_\nu\, p^\nu\, G^\mu\, 
\mathcal{U}\,(M\, M^\dagger_\mu+M_\mu\,M^\dagger)\,
\mathcal{U}^\dagger G-\text{h.c.}\right\}
\end{equation}
where the notation $G^\mu=\partial G/\partial p_\mu$ has been used.

Working out Eq.~(\ref{inicio}) it can be cast as,
\begin{align}
\label{inicio2}
\Delta S_\chi^{RR}=&\frac{1}{4}\int \frac{d^4 p}{(2\pi)^4}\,\,
e^{i\,p\cdot r}
\left\{\sigma_\mu\left(
\begin{array}{cc}
G_1^{\,2}\, \mathcal{A}_{11}^\mu & G_1 G_2\,\mathcal{A}_{12}^\mu\\
-G_1 G_2\,\mathcal{A}_{12}^{\mu\,*} & -G_2^{\,2} \mathcal{A}_{11}^\mu
\end{array}
\right)\right.\nonumber\\
& \nonumber\\
+&\left. (\sigma\cdot p) \left(G_{1\, \mu} G_2-G_1 G_{2\, \mu}\right)
\left(\begin{array}{cc}
0& \mathcal{B}_{12}^{\,\mu}\\
-\mathcal{B}_{12}^{\mu\,*}&0
\end{array}
\right)
\right\}
\end{align}
The calculation of the coefficients $\mathcal{A}_{11}^\mu$, 
$\mathcal{A}_{12}^\mu$ and $\mathcal{B}_{12}^{\,\mu}$ is straightforward.
We will only give the relevant pieces that contribute to the CP violating
current (\ref{cor}).
\begin{align}
\label{coeficientes}
\text{Im} \left[\mathcal{A}_{11}^\mu\right]=&
-\frac{\text{Im}(M_2\mu)}{\Lambda}\left( u_1 u_2^\mu+u_2
u_1^\mu\right)\nonumber\\ \text{Im} \left[\left(M_2 u_1+\mu\,
u_2\right)\mathcal{A}_{12}^\mu\right]=&
-\Delta\frac{\text{Im}(M_2\mu)}{\Lambda}\left( u_1 u_2^\mu+u_2
u_1^\mu\right)\nonumber\\ \text{Im}\left[ \left(M_2 u_1+\mu\,
u_2\right)\mathcal{B}_{12}^\mu\right] =&-\
\text{Im}(M_2\mu) \left( u_1 u_2^\mu-u_2 u_1^\mu\right)
\end{align}

We can see from (\ref{coeficientes}) that the combination $u_1
u_2^\mu-u_2 u_1^\mu$ appears in the off-diagonal coefficients of
$\Delta S_\chi^{RR}$. Had we neglected the latter we would have
obtained only the orthogonal combination $u_1 u_2^\mu+u_2 u_1^\mu$ as
claimed in Ref.~\cite{plus}.  However, contrary to the claim made by
these authors, we do not find that the off-diagonal terms lead to a
contribution of second-order to the currents relevant for electroweak
baryogenesis. Since this is a crucial difference between our results
and those obtained in Refs.~\cite{plus,Kainulainen:2001cn}, we will be
more explicit about the origin of the discrepancy with those authors
at the end of this section.

Returning to the derivation of the currents, the last step would be
going from the Green function (\ref{inicio2}) to the CP-violating
current as defined in (\ref{cor2}). To this end we must rotate to the
weak eigenstate basis and project onto the Higgsino component. We must
then define the Green function
\begin{equation}
\label{sb}
S_{RR}^{(B)}=\mathcal{U}^\dagger \Delta S_\chi^{RR}\mathcal{U}
\end{equation}
and the CP-violating background current (\ref{cor2}) is now given by
\begin{equation}
\label{cor3}
j^{(B)\,\mu}_{RR}=\lim_{r\to 0}\Tr\mathcal{P}_{22} 
\sigma^\mu S^{(B)}_{RR}(z+r/2,z-r/2)
\end{equation}
where $\mathcal{P}_{22} = (\sigma_0 - \sigma_3)/2$ with $\sigma_i$ being
the two by two Pauli matrices in flavor space
and $\sigma_0$ being the two by two identity matrix. 
The current $j^{(B)\,\mu}_{RR}$
can be given by the expression
\begin{align}
\label{cor4}
j^{(B)\,\mu}_{RR}=&\frac{1}{4\Lambda}\int \frac{d^4 p}{(2\pi)^4}
\left\{\left[
(\Lambda-\Delta)\,G_1^2-(\Lambda+\Delta)\,G_2^2
\right]\,\text{Im}\,\mathcal{A}_{11}^\mu \right.\nonumber\\
+&
(G_1 G_2+G_2 G_1)\ \text{Im} 
\left[\left(M_2 u_1+\mu\, u_2\right)\mathcal{A}_{12}^\mu\right]
\nonumber\\
+&\left. 2\,p^\mu\left(G_{1\,\nu} G_2-G_1 G_{2\,\nu}\right)\
\text{Im}\left[ \left(M_2 u_1+\mu\, u_2\right)\mathcal{B}_{12}^{\,\nu}\right]
 \phantom{\frac{1}{1}}\right\}
\end{align}
where the coefficients $\mathcal{A}_{11}$, $\mathcal{A}_{12}$
and $\mathcal{B}_{12}$ of the Green function in (\ref{coeficientes}) 
are made explicit. The final expression can be written as,
\begin{align}
\label{cor5}
j^{(B)\,\mu}_{RR}=&-\frac{\text{Im}(M_2\mu)}{4\Lambda}
\int \frac{d^4 p}{(2\pi)^4}
\left\{\left[
G_1^2 -G_2^2
-\frac{\Delta}{\Lambda}(G_1-G_2)^2
\right]
(u_1 u_2^\mu+ u_2 u_1^\mu)
\right.\nonumber\\ & \nonumber\\
+&\left. 2\,p^\mu \
\left( G_{1\,\nu} G_2-G_1  G_{2\,\nu}\right)\ 
(u_1 u_2^\nu- u_2 u_1^\nu)\right\}
\end{align}

The left-handed current $j^{(B)\,\mu}_{LL}$ can be obtained in the
same way.  The corresponding result can be read off from (\ref{cor5})
after making the changes $\Delta\to\bar\Delta$ and $u_1\leftrightarrow
u_2$. From them the vector and axial currents can be computed and the
result obtained from (\ref{cor5}) agrees with that presented in
Eqs.~(3.14) and (3.15) of Ref.~\cite{wefive}.

To conclude this section we would like to make some final comments
concerning the origin of our discrepancy with the results presented in
Ref.~\cite{Kainulainen:2001cn}.  In Ref.~\cite{Kainulainen:2001cn} it
is correctly argued that, in the absence of particle-changing
interactions and in the mass eigenstate basis at a given point $z$,
terms proportional to $u_1 u_2^{\mu} - u_2 u_1^{\mu}$ only appear in
the off-diagonal elements of the two-by-two matrix of Green functions.
This is in agreement with the result obtained in Eq.~(\ref{inicio2}).
From here, the authors of Ref.~\cite{Kainulainen:2001cn} conclude that
the sources receive no contribution proportional to $u_1 u_2^{\mu} -
u_2 u_1^{\mu}$ at first order in the derivative expansion.  This
conclusion is not correct as we will now explain.

First of all, let us emphasize that the particle changing interactions
are essential in order to convert the original Higgsino density into a
left-handed quark one. In the absence of these interactions with the
plasma the net result for the baryon asymmetry will be much smaller
than the one required for consistency with Big-Bang
Nucleosynthesis. The Higgsino states can not be associated with the
mass eigenstates at any point $z$ along the bubble wall, where the
Higgs background is non-vanishing. In order to treat the particle
changing interactions one has to transform to the weak interaction
basis.

Second, observe that the currents are completely determined by $\Delta
S_\chi^{RR}$ ($\Delta S_\chi^{LL}$).  The dominant, diagonal
contribution $S^{RR}$ ($S^{LL}$) plays no role in the definition of
the CP-violating densities at this order. The CP-violating densities
receive contributions from both the diagonal and off-diagonal terms
after rotating to the weak eigenstate basis as in (\ref{sb}) and both
contributions appear at first order. Therefore, one should compute
both the diagonal and off-diagonal contribution to the Green
functions.  The fact that the diagonal terms have no dependence on the
combination $u_2 u_1^{\mu} - u_1 u_2^{\mu}$ does not imply that the
sources share such an independence, since the off-diagonal terms in
the mass eigenstate basis play an important role when interactions are
included. Actually, as shown in Ref.~\cite{wefive}, depending on the
parameter space, the contributions proportional to $u_1 u_2^\mu-u_2
u_1^\mu$ may be the dominant ones in the generation of a relevant
baryon asymmetry of the universe.

Similar arguments can be used to show that the considerations of the
authors of Ref.~\cite{plus}, leading to the absence of a dependence of
the sources on $u_2 u_1^{\mu} - u_1 u_2^{\mu}$ can not be sustained.
Indeed, Ref.~\cite{Kainulainen:2001cn} tries to provide a formal proof
to these considerations. Moreover, the authors of
Ref.~\cite{Kainulainen:2001cn} neglect the chargino mixing in the
interaction terms, in spite of the fact that mixing is essential for a
non-vanishing source. Therefore, their treatment relies on an
approximation that is invalid in the interesting region where $M_2$ is
close to $\mu$. As we showed in Ref.~\cite{wefive} it is precisely in
this region of parameters where the sources proportional to $u_2
u_1^{\mu} - u_1 u_2^{\mu}$ become most relevant in the baryon
asymmetry generation (see discussion below).

\section{\sc Implications for the baryon asymmetry}

The detailed evaluation of the currents necessary to compute the
sources for the diffusion equations presented in the last section was
already provided in a previous article~\cite{wefive}. In
Ref.~\cite{wefive}, the resulting baryon asymmetry was computed under
the assumption that the connection between the sources for the
diffusion equations and the CP-violating currents was that presented
in Refs.~\cite{Toni2,hn}, namely $S_i = \Gamma^T_i n_i^{(B)}$. In
section 2 we showed that, contrary to this assumption, the real
sources, Eq.~(\ref{source}), have a form that depends on derivatives
of the CP-violating currents computed in the presence of the
background fields. Therefore, the resulting value of the baryon
asymmetry is different from the one computed in Ref.~\cite{wefive}.

The baryon asymmetry resulting from the solution of the diffusion
equations depends on integral functions along the wall of the
temperature dependent value of the sources times exponential factors
which are slowly varying along the bubble
wall~\cite{wefive}. Therefore, the new values can be simply obtained
from the earlier ones by integration by parts, taking care of the
proper boundary conditions.

As discussed in the introduction, the total baryon asymmetry depends
on two different contributions.  The first one is proportional to
\begin{equation}
\epsilon_{ij} H_i \partial_{\mu} H_j = v^2(T) \partial_{\mu}(\beta).
\label{asymm}
\end{equation}
This expression, Eq.~(\ref{asymm}), is proportional to the variation
of the angle
$$\beta = \arctan\left[v_2(T)/v_1(T)\right]$$ 
at the wall of the expanding bubble, which tends to zero for
large values of $m_A$. Furthermore, independently of $m_A$, for large
values of $\tan\beta$, $\beta$ varies only slightly from its value
$\beta \simeq \pi/2$ and therefore the baryon asymmetry tends to be
suppressed. Finally, this contribution to the sources of the baryon
asymmetry has a resonant behaviour for $M_2 = |\mu|$. Therefore, it
becomes more relevant for moderate values of $m_A$ and $\tan\beta$ and
for values of $M_2 \simeq |\mu|$.

The second contribution to the baryon asymmetry depends on 
\begin{equation}
H_1 \partial_{\mu} H_2 + H_2 \partial_{\mu} H_1 = 
v^2 \cos(2\beta) \partial_{\mu} \beta \; + \;
v \partial_{\mu} v \sin(2\beta)  
\label{symm}
\end{equation}
Similarly to (\ref{asymm}), the first term in the expresion
(\ref{symm}) is suppressed for large values of $m_A$ and/or
$\tan\beta$.  Although the second term is also suppressed for large
values of $\tan\beta$, it is unsuppressed for large values of $m_A$.
Therefore, we expect this contribution to become the dominant one for
large values of $m_A$, particulary in the non-resonant regions with
values of $M_2$ very different from $|\mu|$.

Finally, let us remark that the final baryon asymmetry is obtained by
solving the diffusion equation for the total baryon number.  This
equation may be easily derived by summing up the diffusion equations
for the thirty six chiral quarks of the Standard Model (three
generations, two flavors of quark per generation, two chiralities and
three colors per quark). The sum of the quarks densities is just three
times the total baryon number density (each quark carries a baryon
number $1/3$).  The variation of the baryon number is governed by the
sphaleron processes, which affect only the left-handed chiral quarks
and leptons.

The left-handed quark chemical potentials receive two
contributions. The dominant one may be obtained by solving the
diffusion equation for the different colors of quarks in the presence
of gauge, Yukawa, mass and strong sphaleron interactions. Since no
baryon number violating processes are included, the solutions to these
diffusion equations lead to an equal number of baryons of a given
chirality and antibaryons of the opposite chirality.  These dominant
densities may be considered as approximately constant during the
characteristic long times in which the weak sphaleron processes take
place.

The left-handed densities receive also a subdominant contribution
coming from the weak sphaleron interactions. This contribution is
associated with a net-baryon number which, considering effective
mixing between the different flavors and colors of quarks, is shared
in approximately equal parts by all 36 of them.  Finally, there is
also a net lepton number created which due to charge conservation is
equal to the net baryon number.

Taking all these considerations into account, and considering the
right-handed leptons out of equilibrium, one arrives to the equation
\begin{equation}
D n_B''(z)  - v_\omega  n_B'(z)  =  \theta(-z)
\Gamma_{ws} \left( \frac{3 \; T^2 \; \mu_L^{\rm diff}(z)}{4}
+ A n_B(z) \right)
\end{equation}
where $\Gamma_{ws} = 6 k_{ws} \alpha_w^5 T$ ($\alpha_w$ is the weak
coupling constant), with $k_{ws} \simeq 20$~\cite{sphalT} being the
weak sphaleron rate, while $\mu_L^{\rm diff}$ is equal to the sum over
the three generations of the left-handed up and down quark chemical
potentials associated with a given color of quarks, as obtained by
solving their diffusion equations in the absence of the slow sphaleron
interactions.  In the numerical simulations, we have considered heavy
all squarks and sleptons except for the lightest stop, in which case
the coefficient $A = 24/7$, which differs only slightly from the SM
result $A^{SM} = 15/4$ (see also Ref.~\cite{plus}).

\begin{figure}[htb]
\vspace{.8cm}
\centering
\epsfig{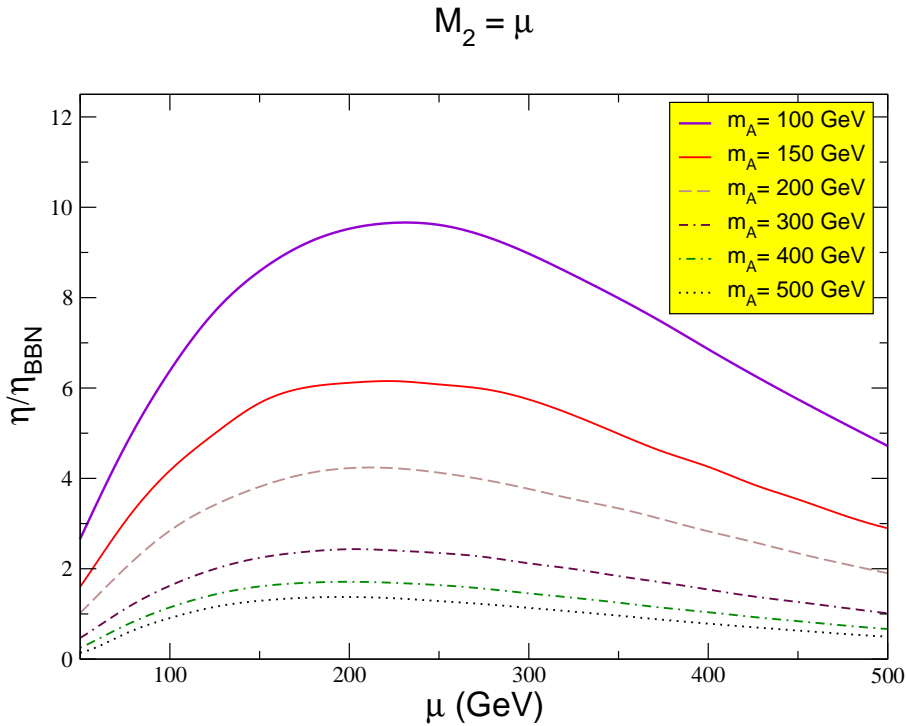}\vspace{.2cm}
\caption{\it Plot of $\eta/\eta_{BBN}$ as a function of $\mu$ for
$\tan\beta = 10$,
$M_2=\mu$ and the specified values of $m_A$.}
\label{figure1}
\end{figure}

In Fig.~\ref{figure1} we show the comparison of the baryon number to
entropy ratio $\eta$ computed within the present model with the mean
value of the one consistent with Big Bang Nucleosynthesis (BBN),
$\eta_{BBN} \simeq (6 \pm 3) \times 10^{-11}$~\cite{BBN}. We have
chosen soft supersymmetry breaking parameters in the stop sector such
that they lead to a value of the Higgs boson mass consistent with the
present experimental constraints, and the lightest stop light enough
so that the phase transition becomes strongly first order, $v(T)/T
\simgt 1$~\cite{marianotalk}.

Since the dominant component of the sources comes from the chargino
sector~\cite{CQRVW,wefive}, the results of the baryon asymmetry depend
only indirectly on the stop parameters, through the values of $v(T)/T$
and $\beta(T)$. Different choices of the parameters lead to variations
of the final result by a factor of order one. Since the method of
computation has implicit uncertainties of similar order, so far
$v(T)/T \simgt 1$, the specific choice of the parameters does not
affect the results for the baryon asymmetry in any significant way.

We have chosen a maximal value of the phase of the $\mu$ parameter,
$\sin \phi_{\mu} = 1$. Therefore, the inverse of the values shown in
the figure can be interpreted as the value of $\sin\phi_{\mu}$
necessary to obtain a prediction consistent with BBN. Observe that due
to the uncertainty in the value of $\eta_{BBN}$~\cite{BBN}, the value
of $\eta_{BBN}$ may be a factor 2 smaller than the above quoted mean
value and therefore the value of the phase may be a factor 2 smaller
than the value obtained by the procedure described above.  Hence, from
Fig.~\ref{figure1} we see that for $m_A$ consistent with the present
experimental constraints, values of $\sin\phi_{\mu} \simgt 0.05$ are
preferred.

\begin{figure}[htb]
\vspace{.8cm}
\centering
\epsfig{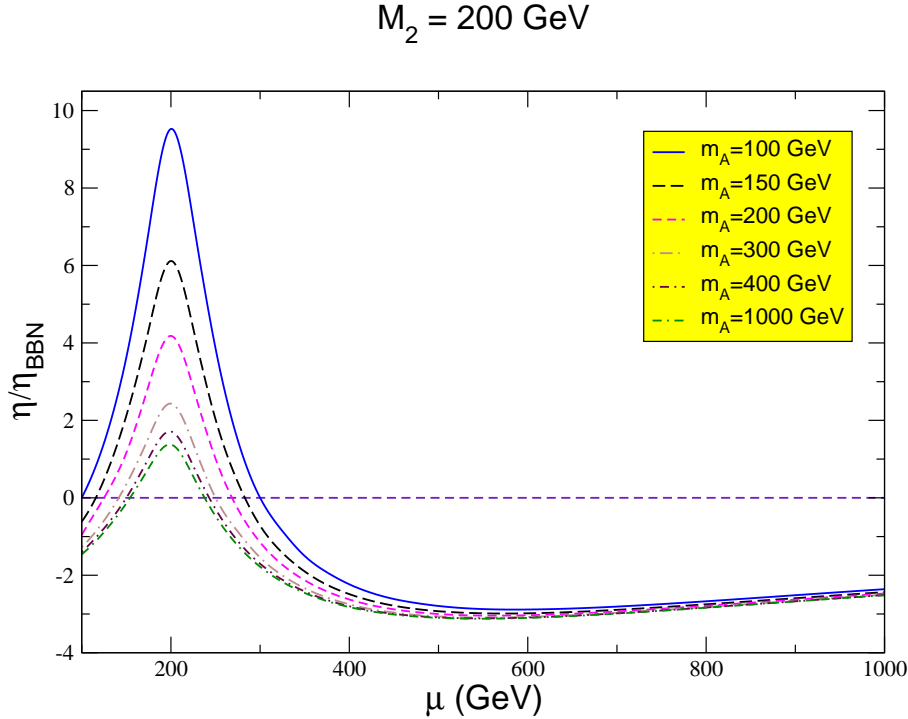}\vspace{.2cm}
\caption{\it Plot of $\eta/\eta_{BBN}$ as a function of $\mu$ for
$M_2=200$ GeV, $\tan\beta = 10$, and the specified values of $m_A$.}
\label{figure2}
\end{figure}

In Fig.~\ref{figure2} we present the values of the ratio
$\eta/\eta_{BBN}$ as a function of $\mu$, for a fixed value of $M_2 =
200$ GeV. For small values of $m_A$ the dominant contribution comes
from the resonant behavior at $M_2 = |\mu|$. However, as $m_A$
increases the maximum of $|\eta|$ is obtained for values of $\mu$ much
larger than $M_2$. The dominant, non-resonant component has opposite
sign to the resonant one and for large values of $\mu$ compared to
$M_2$, it becomes only slightly dependent on $m_A$, in agreement with
our discussion above. We see that even for large values of $m_A$,
values of the baryon asymmetry consistent with BBN predictions may be
obtained for phases of order one. Observe that the baryon asymmetry
tends to zero for large values of $\mu$, since the Higgsino component
of the lightest chargino, present in the plasma, tends to zero in this
limit.

The above results are important in view of the constraints coming from
electric dipole moments~\cite{EDM}. It has been recently observed
that, even in the presence of very heavy squarks and leptons, the
electric dipole moment contributions may not be small, due to the
existence of two loop diagrams involving chargino loops and the non-SM
like Higgs bosons~\cite{recent}.  These contributions are enhanced
with increasing $\tan\beta$, but become rapidly small for large values
of $m_A$ and small mixing in the chargino sector. The results of
Fig.~\ref{figure2} show that a relevant baryon asymmetry may be
obtained even in this particular regime of parameters.

\section{\sc Conclusions}
\label{conclusion}

In this article, we have derived the diffusion equations governing the
result for the baryon density generated by the passage of the
expanding wall at temperatures close to the electroweak phase
transition temperature. Important for this result was the
identification of the sources for the diffusion equations in terms of
the currents generated by the non-trivial Higgs backgrounds, which
vary along the bubble wall. We demonstrated that there are two
different contributions to these sources, one proportional to
$\epsilon^{ij}H_i\partial_{\mu}H_j$ and the other proportional to $H_1
\partial_{\mu}H_2 + (2 \rightarrow 1)$.  Contrary to previous claims,
we showed that none of these contributions to the sources vanish. We
clarified the origin of the discrepancy of our results with those of
Refs.~\cite{plus,Kainulainen:2001cn}.

The two contributions to the sources have a different dependence on
the parameters of the minimal supersymmetric standard model. While the
former contribution has a resonant behaviour for $M_2 = |\mu|$ and
goes to zero for large values of $m_A$, the latter has no resonant
behaviour and remains non-vanishing for large values of $m_A$.

The final result for the baryon asymmetry shows that consistency with
Big Bang Nucleosynthesis demands the relative phase between $\mu$ and
the gaugino mass parameter $M_2$ (or $M_1$, if we consider the
neutralino contribution) to be sizeable. Our calculations lead to a
bound on the value of the CP-violating phase, $\sin\phi \simgt 0.05$,
but due to the natural uncertainties associated with our simplified
treatment of the baryon asymmetry calculation, and the experimental
error in the determination of $\eta_{BBN}$, we cannot reliably rule out
the possibility that somewhat smaller values of the phase $\phi_{\mu}$
may lead to consistency with the BBN predictions.

On the other hand, for sizeable values of the CP-violating phase $\sin
\phi_{\mu}$, cancellation of the potentially large electric dipole
moment contributions is required. These contributions are dominated by
one-loop effects with scalar particles and gauginos participating in
the loop. These corrections may be efficiently suppressed for large
values of the scalar masses of the first and second
generation~\cite{EDM}. Under these conditions, non-vanishing
corrections still appear at the two-loop level~\cite{recent}.  The
most relevant two-loop corrections are enhanced for large values of
$\tan\beta$, but are suppressed for large values of the CP-odd mass
parameter $m_A$. It is reassuring to observe that even for large
values of the CP-odd mass parameter a baryon asymmetry consistent with
BBN may be obtained.

In summary within the framework of the minimal supersymmetric
extension of the Standard Model, the realization of the electroweak
baryogenesis scenario requires the presence of a light Higgs boson,
with mass smaller than about 120 GeV, and a light stop, with mass
smaller than the top quark mass~\cite{mariano1}-\cite{LR},
\cite{marianotalk}. Moreover it has been found that this baryogenesis
mechanism does not guarantee a new CP violating signal at the
$B$-factories~\cite{Murayama:2002xk}.  Therefore, apart from the
experimental constraints coming from electric dipole moments, a
definitive test of the electroweak scenario withih the MSSM will come
from Higgs~\cite{higgs} and stop~\cite{stop} searches at the Tevatron
and at the LHC.

\section*{\sc Acknowledgments}
We acknowledge discussions with T. Prokopec and S. Weinstock. M.Q. and
M.S. would like to thank the Theory Division of the Fermi National
Accelerator Laboratory, where part of this work was done, for
hospitality. M.C. and C.W. also wish to thank the CERN Theory
Division, where part of this work was done, for hospitality.
M.S. would like to thank the Theory Group of Instituto de Estructura
de la Materia, where this work was partly done, for
hospitality. M.Q. would also like to thank the Physics and Astronomy
Department of the Johns Hopkins University for the hospitality
extended to him as Bearden Visiting Professor. The work of C.W. is
supported in part by the US DOE, Div. of HEP, Contract
W-31-109-ENG-38.  Fermilab is operated by Universities Research
Association Inc. under contract no. DE-AC02-76CH02000 with the
DOE. The work of M.S. is supported by the US DOE Contract
DE-A505-89ER40518.

\appendix
\section{\sc The chargino sector}
In this appendix we summarize some useful formulae which concern the
chargino sector of the MSSM.  The chargino mass matrix is given by
\begin{equation}
\label{masach}
M(z)=\left(
\begin{array}{cc}
M_2 & u_2(z) \\
u_1(z) & \mu_c
\end{array}
\right)
\end{equation}
where we have defined $u_i(z)\equiv g H_i(z)$.  The diagonalizing
matrices are
\begin{align}
\label{UV}
\mathcal{U}=&\frac{1}{\sqrt{2\, \Lambda(\Delta+\Lambda)}}
\left(
\begin{array}{cc}
\Delta+\Lambda & M_2\, u_1+\mu^*_c\, u_2 \\
-\left( M_2\, u_1+\mu_c\, u_2 \right) & \Delta+\Lambda
\end{array}
\right)
\nonumber\\
\mathcal{V}=&\frac{1}{\sqrt{2\, \Lambda(\bar\Delta+\Lambda)}}
\left(
\begin{array}{cc}
\bar\Delta+\Lambda & M_2\, u_2+\mu_c\, u_1 \\
-\left( M_2\, u_2+\mu^*_c\, u_1 \right) & \bar\Delta+\Lambda
\end{array}
\right)\ ,
\end{align}
where field redefinitions have been made in order to make the Higgs
vacuum expectation values, as well as the weak gaugino mass $M_2$,
real,
\begin{align}
\label{defin}
\Delta=&(M_2^2-|\mu_c|^2-u_1^2+u_2^2)/2 \nonumber\\
\bar\Delta=&(M_2^2-|\mu_c|^2-u_2^2+u_1^2)/2 \nonumber\\
\Lambda=&\left(\Delta^2+\left|M_2\,u_1+\mu^*_c\, u_2 \right|^2\right)^{1/2}\ ,
\end{align}
and the mass eigenvalues are given by
\begin{align}
\label{eigenval}
m_1(z)=& \frac{\left(\Delta+\Lambda+u_1^2(z)\right)M_2+u_1(z) u_2(z)\mu^*_c}
{\sqrt{(\Delta+\Lambda)(\bar\Delta+\Lambda)}}
\nonumber\\
m_2(z)=& \frac{\left(\Delta+\Lambda-u_2^2(z)\right)\mu_c-u_1(z) u_2(z)M_2}
{\sqrt{(\Delta+\Lambda)(\bar\Delta+\Lambda)}} \ .
\end{align}

In the mass eigenstate basis the free fermionic Green functions can be
written in terms of the bosonic ones $G(p;z)$ as
\begin{xalignat}{2}
\label{relations}
S^{RR}(p;z)=&\sigma_\mu p^\mu G(p;z)&\qquad 
S^{RL}(p;z)=&\begin{pmatrix}m_1(z)&0\\0&m_2(z)\end{pmatrix} G(p;z)
\notag\\
S^{LR}(p;z)=&\begin{pmatrix}m^*_1(z)&0\\0&m^*_2(z)\end{pmatrix} G(p;z)
&\qquad S^{LL}(p;z)=&\overline{\sigma}_\mu p^\mu G(p;z)
\end{xalignat}
and
\begin{equation}
\label{scalar}
G=\left(
\begin{array}{cc}
G_1(p;z) & 0\\
0 & G_2(p;z)
\end{array}
\right)
\end{equation}
where $G_i$ is the two-by-two matrix of a bosonic propagator
corresponding to the mass $|m_i(z)|$ (\ref{eigenval}) given by

\begin{eqnarray}
\label{prop}
G_i^{11}&=&P_i^+ -f_F \left(P_i^+ -P_i^-\right)\nonumber\\
G_i^{12}&=&\left[\theta(p^0)-f_F\right] \left(P_i^+ -P_i^-\right)\nonumber\\
G_i^{21}&=&\left[\theta(-p^0)-f_F\right] \left(P_i^+ -P_i^-\right)\nonumber\\
G_i^{22}&=&-P_i^- -f_F \left(P_i^+ -P_i^-\right)\ ,
\end{eqnarray}
where $f_F\equiv n_F(|p^0|)$ is the Fermi-Dirac distribution function
in equilibrium, which contains the dependence on the temperature $T$,
\begin{equation}
\label{pes}
P_i^{\pm}=\frac{1}{p_0^2-\vec{p}^2-|m_i(z)|^2\pm 2 i\Gamma_{i}|p^0|}
\ ,
\end{equation}
and $\Gamma_{i}$ is the particle width.

\end{document}